# AI-Induced Human Responsibility (AIHR) in AI–human teams

Greg Nyilasy[a] (corresponding author)
Brock Bastian[b]
Jennifer Overbeck[c]
Abraham Ryan Ade Putra Hito[d]

[a, d]Department of Management and Marketing, Faculty of Business and Economics, the University of Melbourne, 198 Berkeley St, Carlton, VIC 3053, Australia
[b]School of Psychological Sciences, Faculty of Medicine, Dentistry, and Health Sciences, the University of Melbourne, Parkville, VIC 3010, Australia
[c]Melbourne Business School, the University of Melbourne, 200 Leicester St, Carlton, VIC 3053, Australia

**Statements and Declarations**

**This study was funded by the Australian Government Research Training Program and the University of Melbourne's Doctoral Program in Marketing scholarship.**

**The authors have no competing interests to declare that are relevant to the content of this article.**

# AI-Induced Human Responsibility (AIHR) in AI–human teams

1. Introduction

Imagine this: An internal audit flags that a bank's "AI-assisted" lending team – which paired an AI model with a human employee – has been systematically rejecting qualified applicants from a legally-protected demographic group. The loan officer insists she followed the workflow; the vendor argues the model merely "recommended" outcomes; the compliance unit can't isolate a single decisive action that caused the harm. Yet the organization still faces a deceptively simple question: *Who is responsible when a morally consequential mistake emerges from a human–AI team, and the causal chain is ambiguous?* Intuition suggests that the algorithm is an easy target, especially when public skepticism toward AI is already high, and that the human teammate will be motivated to deflect blame.

We argue, however, that the psychology of moral responsibility allocation may run in the opposite direction. Inspired by changes in how humans and AI are capable of interacting with the introduction of novel AI technologies such as Large Language Models (LLMs) (Ratliff, 2026), as well as related research on human-AI teaming (Schmutz et al., 2024), the research reported in this paper examines whether moral responsibility may be more likely assigned to the *human* member of a human-AI team, even if that means assigning responsibility to oneself. We propose and test an effect that we call AI-Induced Human Responsibility (AIHR).

An AIHR effect is counterintuitive for at least two key reasons. First, AI algorithms have an organizational trust deficit (Li & Bitterly, 2024), especially when moral risks are apparent (Dietvorst & Bartels, 2022; Hillebrand et al., 2025). Studies have confirmed that humans readily blame AI when something goes wrong (Bigman et al., 2019; Dong & Bocian, 2024). A tendency to blame AIs is also evident in the marketplace, such as when Elon Musk's AI company xAI faced outrage in mid-2025 because its chatbot Grok began issuing vile, extremist responses on X/Twitter (Taylor, 2025). Such moral risks add to business organizations' hesitation to fully embrace the technology for sensitive tasks (Chakravorti, 2024).

Second, people are generally rather good at shifting blame away from themselves, in line with the well-documented self-serving bias (Larson, 1977; Staw et al., 1983). Thus, when objective causes for organizational mishaps are ambiguous, one would predict shifting blame to any suitable target (especially if that target, such as AI, is believed to be untrustworthy anyway). Therefore, in morally charged organizational situations with AI involvement, one may expect human moral disengagement (Ogunfowora et al., 2022), rather than increased human moral responsibility.

However, AI-teaming (incorporating AI algorithms into human work teams; Schmutz et al., 2024) presents a different reality. Employees are increasingly encouraged to pair up with AI for various organizational decisions, including those with moral consequences (Dell'Acqua et al., 2025) – something that has only recently been made possible by the widespread adoption of Large Language Models (Zehnle et al., 2025). LLMs' realistic interactive capabilities have made AI-teaming not simply a metaphor, but reality, mimicking human-human team interaction. This realism has also made it possible for AI-human teams to be involved in decisions with immediate moral implications, such as job application assessments, lending approvals, and insurance claim decisions (Chen, 2023).

Further, AI algorithms in such paired settings, by definition, operate *jointly* with humans rather than in isolation, which may change how humans perceive AI (Raisch & Fomina, 2025).

In such mixed responsibility scenarios, errors or mistakes are plausibly results from *joint* failures of humans and AI and thus are expected to be judged so by human observers. Given the closer involvement between the teammates in a human-AI team interaction, it is less likely that broad negative prior evaluations of AI ("algorithm aversion"; Schilke & Reimann, 2025) will suffice for forming an appropriate moral judgment. As we will demonstrate, a more considered and deeper set of inferences about the likely lack of true AI autonomy (and, instead, affirmed human autonomy) may cause humans to shift blame to themselves when a mistake occurs.

This paper makes three contributions to the study of AI deployment in teams. First, we document a new effect that we label AI-Induced Human Responsibility (AIHR), complementing earlier work on algorithm aversion positing that people readily assign blame to AI algorithms after an error (Dietvorst et al., 2015) or blame AI more for the same error (Bonnefon et al., 2024). We find that, under conditions of joint moral responsibility between two team members, people shift responsibility onto *humans* rather than AI in a human-AI team. This is despite the fact that both the human-AI and control human-human teams were described as working on identical tasks and the scenarios provided no apparent reasons to shift blame this way (or indeed in any other way than randomly).

Second, we demonstrate that the effect is robust enough to counteract a well-established and large social-psychological effect, the self-serving bias (Staw et al., 1983). Even when respondents had to assign responsibility for an error to *themselves* to avoid blaming the AI, they did so. This is even more surprising given that, again, they had no reason to believe they were responsible for the error, and indeed, they did nothing wrong. AIHR is strong enough to make people internalize responsibility under ambiguity (or, arguably, even against the normatively rational certainty that they were *not* responsible). This effect trumps even social desirability, given the socially undesirable nature of the error in some scenarios.

Third, we pinpoint a mechanism, perceived autonomy (Bigman et al., 2019), which AI is believed to lack in morally ambiguous joint responsibility scenarios. Doing so, we also complement prevailing explanations for human assessments of AI, such as mind perception (Gray et al., 2007). We demonstrate that it is specifically perceived autonomy and not AI mind perception that mediates our effect. We suggest that, within organizational settings, perceptions of autonomy extend beyond perceptions of mental abilities (e.g., Waytz, Gray, et al., 2010) to include inferences based on social norms around AI use and structural factors such as expected roles and responsibilities within teams (see Humphrey et al., 2007; Seers, 1989). In the case of our theory of AIHR, perceived responsibility of AI is shifted onto humans not because an AI is viewed as lacking the mental ability to make such moral decisions, but rather because of expectations derived from social norms and structural constraints in its usage within these contexts, limiting its perceived autonomy.

## 2. The Theory of AI-Induced Human Responsibility (AIHR)

*2.1. The backdrop of algorithm aversion and self-serving biases in organizations*

Organizations are rapidly moving from "humans use AI tools" to human–AI teaming, or dyadic or team-based arrangements in which AI is embedded in the workflow and jointly shapes decisions and outcomes (Raisch & Fomina, 2025; Schmutz et al., 2024). This shift matters for moral responsibility because many organizational mistakes are not purely technical – they produce moral harm (e.g., discriminatory screening, unfair lending, wrongful denial of insurance claims) in contexts where causality and intent are ambiguous (Alicke, 2000; Malle et al., 2014)

In terms of finding the culprit and assigning responsibility between human and AI, one may predict for at least two reasons that people would blame the AI more than the human. First, empirical evidence suggests that humans are indeed ready to "blame the bot" when something goes wrong (Crolic et al., 2022; Dietvorst et al., 2015; Gill, 2020). This is not surprising given how negatively predisposed humans are towards AI, even when it functions appropriately (Burton et al., 2020; De Freitas et al., 2023). Aversion to AI only meaningfully diminishes (and may turn into algorithm appreciation) when task fit and positive utility are very clearly demonstrated (Logg et al., 2019; Qin et al., in press). However, task fit and positive utility are not likely to be on anyone's mind when a morally harmful error has just occurred – resulting in blaming the AI.

Second, humans may be motivated to evade responsibility due to self-serving bias (Darke & Chaiken, 2005; Larson, 1977). Organizational settings, in general, provide ample opportunity for self-serving behaviors (Staw et al., 1983). But the presence of AI may specifically invite such behaviors through the concept of *responsibility gaps*, the perceived impossibility of assigning responsibility due to AI's opaque nature (Matthias, 2004) and *responsibility abundance*, the impossibility of clear attribution due to the presence of multiple possible moral agents (Kiener, 2025). Humans may be motivated, given the tendency to self-serve, to take advantage of a responsibility gap and avoid responsibility for moral harm.

Yet, these expectations are built largely on settings that treat the human and the algorithm as separable – or that treat the human as a downstream consumer of an AI output (Benk et al., 2025). AI–human teaming differs (Schmutz et al., 2024): It places human and AI inside a shared action system in which each is framed as a "team member," and responsibility must be partitioned within that joint system rather than assigned to a single isolated actor (Choudhary et al., 2025). *Shell Game*, a recent, award-winning podcast series about a human startup founders' work relationships with AI co-founders, documents these changed psychological relationships between human and machine with startling realism (Ratliff, 2026).

*2.2. AI-Induced Human Responsibility (AIHR)*

We propose that AI-human teaming can *increase* (not decrease) perceived human moral responsibility after a morally consequential mistake under ambiguity. The core reason is that teamwork invites a different attributional analysis than individual decision-making.

Whereas prior work investigating humans' reactions towards AI or human moral decision-making treated the parties as operating in isolation (Dietvorst & Bartels, 2022), the integration of AI into work teams represents a new reality. Team contexts prompt observers to explain outcomes using not only attitudinal ("algorithm aversion") or cognitive capacity ("mind perception") accounts, but also a more discerning analysis of structural constraints and social norms around these constraints in the workplace (Humphrey et al., 2007).

The deployment of AI–human teams suggests that participants are acting conjointly rather than in an individually isolated manner (Choudhary et al., 2025; Raisch & Fomina, 2025). As human and AI actions intertwine, moral judgment on allocating responsibility is naturally more complex than individualistic judgments. Following a morally consequential mistake, the presence of multiple actors necessitates apportioning moral responsibility among the involved actors (Malle et al., 2022; Naquin & Kurtzberg, 2004). In ambiguous situations, especially, where such partitioning of accountability cannot be achieved in an unequivocal, heuristic manner, human observers engage in a more involved moral calculus to determine who is responsible (Malle et al., 2014).

Second, as part of this deeper processing, the complexity and opacity of team settings prompt human actors or observers to consider structural constraints and social norms around these constraints that go beyond simplistic "blame the bot" defaults (Crolic et al., 2022; Dietvorst et al., 2015). For example, observers may consider the social rules of engagement between AI and human (in which analysis humans are likely perceived to be in control of the AI; Ullman & Malle, 2017). Respondents are likely to make structural inferences: assessing work design and organizational structures, in which (currently, at least) humans maintain a dominant role (Tsamados et al., 2025). Moral responsibility judgments are acutely sensitive to constraints: When an actor's options are restricted by coercion, orders, or narrow role mandates, blame is mitigated because the actor is seen as lacking the control to do otherwise (Alicke, 2000; Cushman, 2008). AI–human teaming makes constraints particularly salient because AI systems are understood to operate within organizational guardrails, instructions, and narrowly specified objectives, even when outputs appear sophisticated (Institute_for_Family_Studies, 2025). Consequently, in an AI–human setting, this type of moral calculus will lead to a responsibility shift to the human, whereas in a human–human team no shift is expected to occur, absent specific information to incriminate any one individual.

In summary, these normative and structural expectations around AI use make a simple "blame the bot" heuristic less psychologically sufficient than our proposed alternatives. When the system is explicitly a team, observers must still identify the locus of discretion and the bearer of role obligations. We argue that under ambiguity – absent explicit cues that either party holds responsibility – observers ascribe accountable moral agency to the human, because humans are the prototypical bearers of moral agency and workplace obligations, whereas machines are perceived as constrained implementers (despite their growing capabilities). This produces AI-Induced Human Responsibility (AIHR):

> **H1**: In situations of organizational moral harm under ambiguity, people internalize responsibility more in AI–human than in human–human work teams.

*2.3. The mediating role of perceived autonomy*

Our proposed mechanism is perceived autonomy. After moral harm occurs, responsibility allocation hinges on whether an actor is perceived to have exercised control over the action and its consequences, what moral psychology variously describes as culpable control, reasons-responsiveness, or the capacity for intentional choice (Alicke, 2000; Cushman, 2008; Malle & Robbins, 2025). A critical component for assigning moral responsibility, therefore, is the perception of the actors' independence in their actions; that is, their degree of autonomy (Malle et al., 2014). When an actor is seen as more autonomous, they are more likely to be held accountable for their actions (Bigman et al., 2019; Vanneste & Puranam, 2025). For instance, an employee who executes a business strategy that later harms some stakeholders would bear less perceived moral responsibility when narrowly following orders compared with when acting independently (Malle, 2021).

Assessing an agent's autonomy informs people of the agent's situational constraints (Bigman et al., 2019). Responsibility is attributed more to an action that stems from the exercise of free will, the authentic ability to make choices among different options, relatively free of external constraints (Feldman, 2017; Monroe et al., 2016). An action resulting from the exercise of free will is considered the product of an entity's conscious deliberation rather than being predetermined or influenced by external pressures. As a consequence of free will, an actor is perceived to have reflected on their values and reasoning before selecting among alternative actions, rather than being mechanically determined (Shariff et al., 2014). For example, in Volkswagen's "Dieselgate" scandal, the mechanics were not held accountable as much as the firm's executives, given that these mechanics committed unethical actions due to pressure from the executives (Hotten, 2015).

An AI can be said to possess autonomy, given its ability to take an action independently. However, AI's freedom to choose actions is ultimately constrained compared with that of humans. AI operates within fixed boundaries dictated by its human-written nature, human-constrained training data sources, and operative "guardrails," all of which limit its ability to act outside these boundaries, diminishing its actual – and thus perceived – autonomy (Shariff et al., 2014).

Despite the human tendency for anthropomorphizing technology (Blut et al., 2021; Waytz, Morewedge, et al., 2010), users seem to be aware that AI lacks humanlike autonomy (Astobiza, 2024). Indeed, people perceive AI as lacking the ability to understand the nuances of situations, with its actions aimed at maximizing statistical outcomes instead of being aware of the moral concerns present in a situation (Dietvorst & Bartels, 2022; Longoni et al., 2019).

These AI-autonomy inferences have a predictable consequence for moral responsibility partitioning under ambiguity. The more the AI is seen as a constrained implementer (low autonomy), the more likely the human teammate is, by contrast, seen as the bearer of discretionary judgment: the one who can question, override, contextualize, and take ownership. In other words, responsibility shifts onto humans not because AI is perceived as lacking the capacity to understand moral considerations (a pure mind-perception account), but because AI is perceived as constrained by its ascribed role in the team's social structure – limiting its autonomy and therefore its status as a responsible moral agent. Our second hypothesis (H2) encapsulates this mediating mechanism and Figure 2 depicts our complete model.

**H2**: AI-induced human responsibility attributions are mediated by a perception that AI has lower autonomy compared with humans.

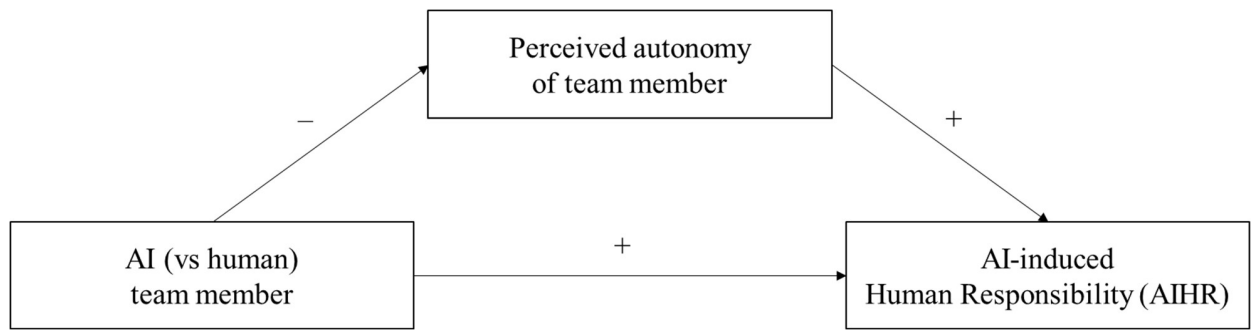

**Fig. 1.** The Theory of AI-Induced Human Responsibility in work teams.

## 3. Overview of studies

*3.1. Overview*

Across four studies, we tested the hypotheses that working alongside an AI in a work team instills a greater sense of responsibility within human team members than in traditional human-only teams. The first study tested the main effect of increased human responsibility in the case of AI–human teams. Study 2 tested the perception of autonomy as the mediating mechanism behind the main effect. Study 3 investigated a social identity-based alternative explanation: threats to one's identity at the workplace (Granulo et al., 2019). Lastly, Study 4 demonstrated robustness across self-other differences in attributions and ruled out mind perception as an alternative explanation. Table 1 summarizes the purpose, main findings, and important features of each study.

**Table 1**
Overview of Studies.

| Study | Main purpose | Study setting/task | Sample size | DV measure (Attribution of responsibility) | AIHR effect (AI–Human vs human–human) on 100 pt scale | Mediating mechanism, alternative explanation | Robustness: Perspective (Self vs other) | Robustness: Harm level (high vs low) | Robustness: Covariates |
|---|---|---|---|---|---|---|---|---|---|
| 1 | Main effect | Bank loan processing: discrimination | 302 | Single item (0-100 slider) | +7.00 points (47.81 vs 40.81) $d = .24$ | - | Self | High | AIHR robust to: age, gender, political philosophy, perceived team performance, social desirability, counterbalancing |
| 2 | Mediation: Perceived autonomy + Robustness across high vs low harm | Bank loan processing: falsification/ irresponsible lending vs filing error | 602 | Three sliders ($\alpha$=.83) | **+14.17 points (High harm)** (65.59 vs. 51.42) $d = .71$ **+11.13 points (Low harm)** (64.96 vs. 53.83) $d = .60$ | Perceived autonomy ✓ | Other | Both and low | AIHR robust to: age, gender, counterbalancing |
| 3 | Ruling out alternative explanation: Threat to self-worth | Bank loan processing: irresponsible lending | 296 | Three sliders ($\alpha$=.85) | +10.36 points (58.85 vs 48.49) $d = .52$ | Threat to self-worth (NS) | Self | High | AIHR robust to: age, gender, counterbalancing |
| 4 | Ruling out alternative explanation: Mind perception + Robustness across other vs self perspectives | Bank loan processing: discrimination | 601 | Three sliders ($\alpha$=.87) | **+9.02 points (Other perspect.)** (64.10 vs 55.08) $d = .40$ **+10.34 points (Self perspect.)** (55.64 vs 45.30) $d = .46$ | Mind perception (NS) | Both self and other | High | AIHR robust to: age, gender, counterbalancing |

*3.2. Research transparency statement*

Past organizational behavior research has emphasized that many published studies are underpowered and that sample-size decisions should be justified via explicit inferential goals rather than rules of thumb (Mazen et al., 1987). Accordingly, we determined target sample sizes a priori using a sensitivity/power approach calibrated to adjacent experimental work on human responses to algorithms in organizational behavior (Logg et al., 2019). Because our focal phenomenon is novel, we did not have a precise ex ante estimate of its magnitude; therefore, we powered each study to detect small-to-moderate effects (using the *pwr* package in R, $d = .35$, $\alpha = .05$, power $= .8$, $n = 130$ per cell per two independent group studies, allowing for online sample technical problems: $n = 150$ per cell). At the end of each study, we collected participants' demographic data (e.g., gender, political philosophy, and age). These demographic data, upon further analysis, did not influence the results. Thus, we do not discuss them in this section. Web Appendices A to D contain all stimuli and measures used in each study, as well as supplementary analyses. While the studies were not preregistered, all data and R codes are available at OSF (https://osf.io/jx2qb/overview?view_only=9214eba061f345769834f7faa9d7b424) This study was funded by the [government body blinded for review] and the [university blinded for review]. The authors have no competing interests to declare that are relevant to the content of this article.

4. **Study 1: Main effect**

Study 1 tested the hypothesis that working in a human–AI (vs. a human–human) team leads employees to attribute more responsibility to themselves in the event of an unfavorable outcome. In this study, participants were asked to imagine a scenario where they teamed up with either a human employee or an AI system to assess and approve loan applications.

*4.1. Method*

*4.1.1. Participants and design*

A total of 302 US residents (US, UK, Ireland, Australia, Canada, and New Zealand nationalities, all English native speakers) were recruited from Prolific, an online research panel, to participate in the study. The participants ($M_{age}$ = 39.9 years; $SD_{age}$ = 14.4 years; 49% female, 50% male, 1% other) were randomly assigned to one of the two (AI–Human vs. Human–Human team) between-subjects conditions.

*4.1.2. Experimental procedure*

Participants read stimulus materials in which those in the AI–Human team condition were paired with an AI system and in the human-human condition paired with a colleague. The scenario told respondents that the team they were part of processed loan applications, approved them, and determined the final loan amount. Later, all participants (irrespective of condition) were told that the loan provider had received discriminatory lending complaints and that an internal audit found their team responsible for the mistake.

After reading the stimuli, the participants attributed responsibility between themselves and the team member (i.e., a human colleague or an AI system) on a scale of 0 to 100 using a slider (0 = "I am not responsible," to 100 = "I am responsible"). We randomized the left-right order of the label on the slider to rule out order effects (Lensvelt-Mulders et al., 2005). This randomization occurred in subsequent studies every time a slider was used.

We measured participants' perceptions of team performance as a control (adapted from Aubé & Rousseau, 2005; three items; $\alpha$ = .90), given that being paired with an AI or another human might influence participants' assessment of their team's capabilities. Additionally, we measured participants' tendency to view themselves positively, operationalized with the Marlowe-Crowne

Social Desirability scale (adapted from Ventimiglia & MacDonald, 2012; four items, $\alpha$ = .66) and used as a control variable. Maintaining a positive self-view is a psychological need that might influence participants' evaluations during the study (Epley & Dunning, 2000). At the end of the study, participants responded to demographic questions and a manipulation check question (asking if the other team member was an AI or a human).

*4.2. Results*

A chi-square test of independence showed a significant association between condition and manipulation-check responses, $\chi^2(1, 302) = 260.30, p < .001$. Participants in the AI–Human condition almost always correctly identified the other agent as AI (94%), as did participants in the Human–Human condition (99%), indicating that the manipulation of team type was successful.

Testing the core hypothesis, a *t*-test revealed a significant main effect of team type on participants' attribution of responsibility to themselves, $t(300) = 2.11, p = .036, d = .24$. Specifically, participants in the AI–Human condition attributed more responsibility to themselves ($M_{\text{AI–Human}} = 47.81, SD_{\text{AI–Human}} = 33.18$) than did those in the Human–Human condition ($M_{\text{Human–Human}} = 40.81, SD_{\text{Human–Human}} = 23.56$).

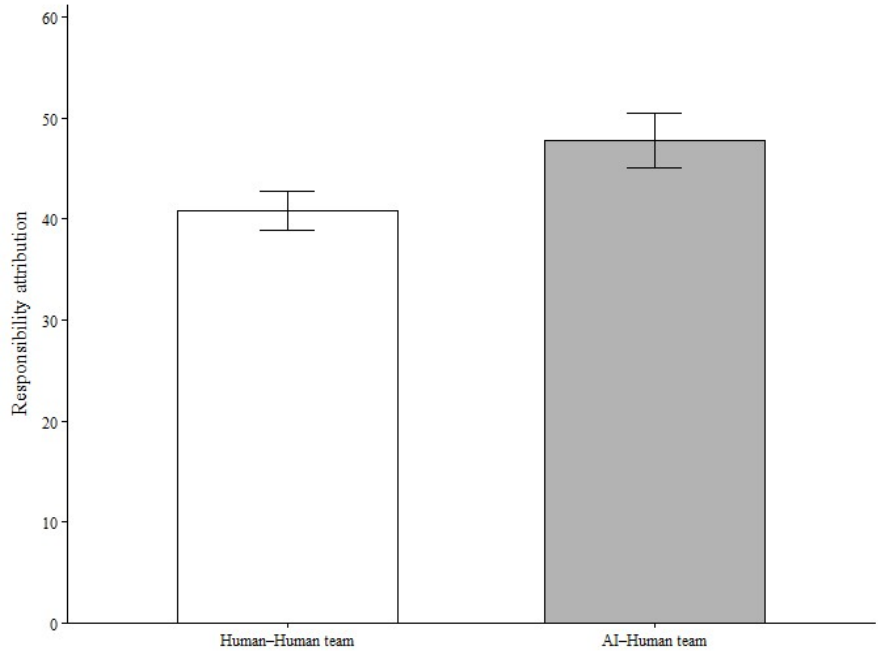

**Fig. 2.** Study 1: The AI-induced Human Responsibility effect: Pairing humans with AI in work teams increases perceived responsibility for the participating humans after moral failure.

For robustness, we investigated the control variables, perceived team performance and social desirability. Perceptions of team performance did not differ by condition, $t(300) = 0.81$, $p = .418$, $d = -.09$; ($M_{AI–Human} = 4.09$, $SD_{AI–Human} = 1.39$ vs. $M_{Human–Human} = 4.22$, $SD_{Human–Human} = 1.39$). We regressed responsibility on team type (contrast coded: $-1$ = Human–Human, $1$ = AI–Human), perceived team performance (mean-centered), and their interaction. Results indicate that higher team performance perceptions were associated with higher human responsibility scores, $b = 5.76$, $SE = 1.15$, $t = 5.01$, $p < .001$. Importantly, however, the main effect of team type remained significant, such that participants in the AI–Human condition reported higher human responsibility scores, $b = 3.88$, $SE = 1.60$, $t = 2.43$, $p = .016$, and the interaction was not significant, $b = 2.24$, $SE = 2.30$, $t = 1.12$, $p = .331$. These findings suggest our focal effect is robust with respect to perceived team performance, which itself is not influenced by our team type manipulation.

Repeating this analysis for the social desirability control variable (positive self-view), we found that social desirability did not differ by condition, $t(300) = -0.17$, $p = .919$, $d = .03$; ($M_{\text{AI–Human}} = 4.88$, $SD_{\text{AI–Human}} = 0.97$ vs. $M_{\text{Human–Human}} = 4.85$, $SD_{\text{Human–Human}} = 1.03$). Then, regressing responsibility on team type (dummy-coded), social desirability (mean-centered), and their interaction, the analysis revealed a non-significant effect of social desirability, $b = 1.84$, $SE = 2.31$, $t = .80$, $p = .426$. Importantly, the main effect of team type remained significant, such that participants in the AI–Human condition reported higher human responsibility scores, $b = 3.50$, $SE = 1.66$, $t = 2.11$, $p = .036$, and the interaction was not significant, $b = -4.01$, $SE = 3.34$, $t = -1.20$, $p = .231$. This serves as evidence that social desirability does not interfere with our results.

*4.3. Discussion*

Study 1's results support our first hypothesis that, when unfavorable moral outcomes occur, individuals attribute more responsibility to a human actor in the case of an AI–Human team than a Human–Human team. This finding is significant because one may expect people to point their fingers at others for a morally relevant mistake, as people may associate taking responsibility with career risk and negative emotions in organizational settings (Kupor et al., 2018). Indeed, people may even be willing to take further steps to avoid responsibility, such as intentionally delegating choices to others to evade accountability completely (Steffel et al., 2016), and are quite ready to blame AI when given a chance (Bonnefon et al., 2016; Gill, 2020), especially in the moral domain (Dietvorst & Bartels, 2022). In light of this, it is noteworthy that the AIHR effect observed in this study was strong enough to counteract people's well-documented tendencies for deflecting blame as well as for moral algorithm aversion.

Though Study 1 provided initial support for our theory, it did not explain the mechanism behind the observed AIHR effect. It also did not account for the fact that moral mistakes exist on a continuum in terms of the harm they cause. Study 2 addresses these issues by investigating perception of autonomy as the mechanism of the effect and offering two scenarios where the error leads to a high vs low harm.

## 5. Study 2: Mediation and robustness across harm levels

Study 2 tested the perception of autonomy as the mediating mechanism explaining the effect of AI-induced human responsibility observed in Study 1. Further, since Study 1 focused on a scenario with serious harm (discrimination based on protected human attributes), we examined whether the AIHR effect generalizes across high (falsification of financial information) and low (filing error leading to more work) harm levels. Companies face a much larger volume of morally consequential errors with a relatively low level of harm than ones with serious harm. Hence, it is essential to see whether the AIHR effect generalizes to these more common situations. Further, we wanted to test whether the effect generalizes to observing others (when the respondent is not an active participant in the team, but rather a bystander).

*5.1. Method*

*5.1.1. Participants and design*

Six hundred and two US residents (US, UK, Ireland, Australia, Canada, and New Zealand nationalities, all English native speakers) were recruited from Prolific, an online research panel platform, to participate in the study. The participants ($M_{age}$ = 37.0 years; $SD_{age}$ = 12.0 years; 49% female, 50% male, 1% other) were then randomly assigned to one of the four conditions in a 2 (Team type: AI–Human team vs. Human–Human team) x 2 (Harm level: High vs. Low) between-subjects design.

*5.1.2. Experimental procedure*

Participants were randomly allocated to either an AI–Human team or a Human–Human team condition. In the AI–Human team condition, participants read that Employee 1 had teamed up with an AI system to process loan applications, whereas in the Human–Human team condition, Employee 1 teamed with Employee 2. Then, participants were told that an internal investigation found that the team had committed an error. In the High (Low) harm condition, the team was found to have falsified clients' financial information (to have assigned wrong file numbers on clients' loan applications) and consequently provided loan amounts beyond the clients' loan serviceability (caused another team to work overtime to rectify the clients' file numbers).

After reading the stimuli, participants attributed responsibility on 0–100 sliders (three items, $\alpha$ = .83, adapted from Hinds et al., 2004; Lei & Rau, 2021) between Employee 1 and the team member on the other side of the slider (either an "Artificial Intelligence system" or "Employee 2"). As in Study 1, the anchor positioning of the sliders was counterbalanced. Participants then assessed the relative perceived autonomy of team members using similar counterbalanced sliders (six items; $\alpha$ = .86; adapted from Bacharach & Aiken, 1976; Berretta et al., 2023). The slider scores were transformed in such a way that higher scores meant more responsibility and more perceived autonomy attributed to Employee 1, who was always a human, whereas the other end of the slider was either Employee 2 or an AI system, depending on the team type condition. Lastly, participants responded to demographic questions and two manipulation check questions (one for each factor). For team type, the categorical manipulation check question asked whether the other team member was an AI or a human. For level of harm, the categorical manipulation check asked whether the harm was falsification of clients' financial information (High harm) or a filing error (Low harm).

*5.2. Results*

To test whether the team type manipulation worked, we regressed the team type manipulation-check item on the experimental factors team type (AI–Human vs. Human–Human team) and harm level (High vs. Low), including their interaction, using logistic regression. IV strongly predicted perceived team type, $b = 6.47$, $SE = .62$, $z = 10.38$, $p < .001$, corresponding to an odds ratio of 647.87, 95% CI [213.60, 2556.56]. In contrast, harm level ($p = .544$) and the team type × harm level interaction ($p = .923$) were not significant, indicating that the harm manipulation did not contaminate the team type manipulation check. Descriptively, participants' manipulation-check responses were consistent with the assigned team type for 96.6% of those in the Human–Human condition and 95.8% of those in the AI–Human condition (misclassification rates 3.4% and 4.2%, respectively). For the harm manipulation, we regressed perceived harm on harm level and team type, plus their interaction. Harm level was a very strong predictor of perceived harm, $b = 4.94$, $SE = 0.31$, $z = 16.08$, $p < .001$, OR = 139.10, 95% CI [78.16, 261.13], whereas team type ($p = .582$) and the harm level × team type interaction ($p = .302$) were not significant. Thus, the harm manipulation check was unaffected by team type. In the Low-harm condition, 93.0% of participants gave responses consistent with low harm (7.0% inconsistent), and in the High-harm condition, 91.1% gave responses consistent with high harm (8.9% inconsistent). Overall, both experimental manipulations were successful and largely independent of each other.

To test the main effects and their possible interaction, we used ANOVA, with team type and harm level as predictors and the average of the responsibility sliders as the dependent variable. Results showed a significant effect of team type, $F(1, 598) = 64.36$, $p < .001$, $p\eta^2 = .10$, but no significant effect of harm level, $F(1, 598) = .84$, $p = .359$, $p\eta^2 \approx .00$, or interaction, $F(1, 598) = .45$, $p = .502$, $p\eta^2 \approx .00$. Planned contrasts showed that participants attributed more responsibility to a human in the AI–Human team than in the Human–Human team scenario, in both the High-harm, $t(598) = 6.16$, $p < .001$ ($M_{AI–Human} = 64.59$, $SD_{AI–Human} = 22.68$ vs. $M_{Human–Human} = 51.42$, $SD_{Human–Human} = 10.79$; $d = .71$) and the Low-harm, $t(598) = 5.19$, $p < .001$ ($M_{AI–Human} = 64.96$, $SD_{AI–Human} = 22.74$ vs. $M_{Human–Human} = 53.83$, $SD_{Human–Human} = 14.82$; $d = .60$) conditions.

A similar pattern emerged for perceived autonomy. An ANOVA with team type and harm level predicting perceived autonomy showed a strong main effect of team type, $F(1, 598) = 185.58$, $p < .001$, $p\eta^2 = .24$, but no significant effect of harm level, $F(1, 598) = 0.08$, $p = .779$, $p\eta^2 \approx .00$, and no interaction, $F(1, 598) = 0.85$, $p = .356$, $p\eta^2 \approx .00$. Planned contrasts indicated that participants perceived the human teammate more autonomous than the AI one in both the High-harm condition, $t(598) = 9.0$, $p < .000$ ($M_{AI–Human} = 69.94$, $SD_{AI–Human} = 20.60$ vs. $M_{Human–Human} = 53.52$, $SD_{Human–Human} = 9.75$; $d = 1.04$) and the Low-harm condition, $t(598) = 10.3$, $p < .000$ ($M_{AI–Human} = 71.50$, $SD_{AI–Human} = 18.70$ vs. $M_{Human–Human} = 52.68$, $SD_{Human–Human} = 11.45$; $d = 1.19$).

To estimate the indirect effect of the team type manipulation on responsibility through perceived autonomy, we used PROCESS Model 4 with 10,000 bootstrapping resamples (version 5.0, Hayes, 2025), including harm type as a covariate in both the mediator and outcome models. Team type significantly predicted perceived autonomy, $b = 17.62$, $SE = 1.29$, $t(599) = 13.62$, $p < .001$, with participants in the AI–Human condition reporting higher perceived autonomy. Harm level did not predict perceived autonomy, $b = -0.39$, $SE = 1.29$, $t(599) = -0.30$, $p = .765$. In turn, perceived autonomy positively predicted responsibility when controlling for team type and harm level, $b = 0.30$, $SE = .05$, $t(598) = 6.55$, $p < .001$; harm level was not a significant predictor of responsibility, $b = -1.25$, $SE = 1.46$, $t(598) = -0.86$, $p = .392$. The total effect of team type on responsibility (controlling for harm level) was significant, $b = 12.15$, $SE = 1.51$, $t(599) = 8.03$, $p < .001$, and was reduced but remained significant when perceived autonomy was included in the model, $b = 6.82$, $SE$

= 1.68, *t*(598) = 4.07, *p* < .001. The indirect effect of team type on responsibility via perceived autonomy was significant, *b* = 5.33, *BootSE* = 1.23, 95% CI [3.06, 7.88], corresponding to a partially standardized indirect effect of 0.27, 95% CI [0.16, 0.40].

*5.3. Discussion*

Replicating the result from Study 1, Study 2 showed that in the AI–Human team scenario, participants attributed more responsibility to the human teammate than in the Human–Human case. Importantly, the results of Study 2 support the prediction (H2) that perception of autonomy is a key mechanism underlying the effect of team arrangement on the attribution of responsibility. Notably, the level of harm (high vs. low) did not significantly affect participants' attribution of responsibility, suggesting that the effect generalizes across harm levels.

Although Study 2 identifies perceived autonomy as a key mechanism behind our effect, the significant direct effect suggests a role for other causal processes. One plausible alternative is that AI in the workplace threatens employees' self-worth and thereby motivates them to take responsibility for errors. The rise of AI and job-replacement narratives can undermine employees' perceived abilities, contributions, and value to the organization and elicit compensatory behaviors such as taking ownership of mistakes (Frese & Keith, 2015; Granulo et al., 2019; Yam et al., 2023).

We believe, however, that in team settings where AI and human are *jointly* responsible for outcomes, self-worth threat is not a likely mechanism. As Study 2 has shown, the team setting makes human autonomy (and the lack of AI autonomy) salient. Autonomy is a basic psychological need (Deci & Ryan, 2000) and its fulfillment protects against self-worth threats. After experiencing heightened autonomy relative to the AI partner, a human should feel affirmed rather than threatened in their self-worth. Nevertheless, because of the intuitive appeal of such an explanation for the team-type effect, Study 3 aimed to rule out the potential alternative mechanism of threat to self-worth.

## 6. Study 3: Ruling out threat to self-worth as an alternative mechanism

*6.1. Method*

*6.1.1. Participants and design*

Two hundred ninety-six US residents (US, UK, Ireland, Australia, Canada, and New Zealand nationalities, all English native speakers) were recruited from Prolific, an online research panel platform, to participate in the study. The participants ($M_{age}$=39.3 years; $SD_{age}$=13.2 years; 49% female, 49% male, 1% other) were randomly assigned to one of the two (i.e., AI–Human vs. a Human–Human team) between-subjects conditions. Like Study 1, Study 3 asked respondents to take a first-person perspective, imagining that they themselves participate in the work teams along with a counterparty.

*6.1.2. Experimental procedure*

Participants read a stimulus text describing themselves working in a bank. Then, those in the AI–Human (Human–human) team condition read that they had been paired with an AI system (a colleague) to evaluate the requested loan amount against relevant documents, such as identity documents and payslips, and decide whether to accept or change the proposed loan amount. Later, participants were told that the bank's internal compliance department subsequently found regulation violations among the loan approvals. In all conditions, participants were told that the internal audit found that it was their team who was responsible for the mistake in evaluating loan applicants' financial information, leading to the provision of loan amounts beyond applicants' ability to pay back.

After reading the stimuli, the participants allocated responsibility between themselves and the team member (i.e., a human employee or an AI system) on a scale of 0 to 100 using a slider similar to those in Studies 1 and 2, on three items ($\alpha$ = .85). Like before, the slider anchors' positions were counterbalanced, with the resulting scores transformed such that a higher score means more responsibility attributed to the self. Next, adapted from Granulo et al. (2019), we assessed perceived threat to self-worth (*self-threat*) on three items ($\alpha$ = .91), using a seven-point Likert scale from 1 to 7 (i.e., 1= "strongly disagree" and 7 = "strongly agree"). At the end of the study, participants responded to demographic questions and a manipulation check (as before, asking whether the other team member was an AI or a human).

*6.2. Results*

A chi-square test of independence showed a significant association between condition and manipulation-check responses, $\chi^2(1, 296) = 253.62, p < .001$. Participants in the AI–Human condition almost always correctly identified the other agent as AI (97%), as did participants in the Human–Human condition (97%), indicating that the manipulation of team type was successful.

A *t*-test revealed a significant effect of team type on participants' attribution of responsibility ($t(294) = 4.45, p < .001, d = .52$). Specifically, participants in AI–Human teams ($M_{\text{AI–Human}}$ = 58.85, $SD_{\text{Human–Human}}$ = 23.90) attributed a higher responsibility to themselves over the outcome than those in Human–Human teams ($M_{\text{Human–Human}}$ = 48.49, $SD_{\text{Human–Human}}$ = 14.87).

To estimate the mediating effect of the team type manipulation on responsibility through self-threat, we used PROCESS Model 4 with 10,000 bootstrapping resamples (version 5.0, Hayes, 2025). Team type did not significantly predict self-threat, $b = 0.24, SE = 0.18, t(294) = 1.34, p = .182$, indicating that participants in the AI versus human teammate condition did not feel different levels of perceived threat to self-worth. Self-threat negatively predicted responsibility when controlling for team type, $b = –2.14, SE = 0.75, t(293) = –2.85, p = .005$, such that higher perceived threat to self-worth was associated with lower self-attributed responsibility. The total effect of team type on responsibility was significant, $b = 10.36$, $SE = 2.32$, $t(294) = 4.45$, $p < .001$, and remained significant and of similar magnitude when self-threat was included in the model, $b = 10.87$, $SE = 2.30$, $t(293) = 4.72$, $p < .001$, indicating no evidence of mediation. Consistently, the indirect effect of team type on responsibility via self-threat was not significant, $b = –0.51$, $BootSE = 0.44$, 95% CI [–1.48, 0.23].

*6.3. Discussion*

Like in previous studies, the results showed that participants attributed more responsibility to a human (themself) in an AI–Human team compared with a human counterparty. Importantly, team type did not significantly affect participants' perception of threats to their self-worth, and mediation analysis further confirmed that the indirect effect of perceived threat to self-worth on participants' attribution of responsibility was nonsignificant. Further, the effect of self-threat on attribution of responsibility was in the *opposite* direction from the team-type effects we are examining. Thus, we rule out perceived threat to self-worth as the mechanism behind the AIHR effect.

In addition, Study 3 confirms the robustness of our effect in the face of humans' self-serving tendencies (Larson, 1977; Miller, 1999). It is remarkable that, just like in Study 1, respondents took a greater share of responsibility onto themselves compared with when they were paired with a human, despite the fact that they had no good reasons to do so (the scenarios were ambiguous). Our explanation is that the mere pairing with AI creates a structural-normative constraint, forcing a re-evaluation of responsibility shifting it towards humans, even when possibly threatening the self. Study 3 confirms that, indeed, self-threat as a mechanism is not in operation (instead, as Study 2 confirms, perceptions of low AI autonomy are). These two findings explain how our effect can take place even when this involves taking blame onto oneself for no good reason – a behavior that is more likely to be resisted in other circumstances.

**7. Study 4: Investigating the self–other attributional gap and ruling out mind perception**

We found confirmation for AIHR in both self and other-focused scenarios – but also we do know that self–other differences in attributions are well-established in general social cognition research (Jones & Nisbett, 1971) and even in AI research specifically (Dong & Bocian, 2024). Therefore, we wanted to directly compare our effect in self- vs other-attribution scenarios, to gain further confirmation that our effect is robust across these two fundamental social cognition domains (Moskowitz, 2005). We expected that, irrespective of respondents' attributional position (self- vs other-focused), AIHR would hold. Building on our three studies' findings, we argue that teaming AIs with humans activates a sense of autonomy that buffers any perceived threat to the self and leads to a responsibility shift towards humans. Although responsibility for a moral transgression is generally experienced as threatening to the self, we argue that the salience of autonomy, and the awareness of structural constraints limiting the AI's autonomy and responsibility, should be present regardless of whether the respondent is taking a first-person perspective or a threat-neutral third-person (other) perspective. That is, in both conditions, the AIHR should be observed.

Another goal of Study 4 was to rule out mind perception as an alternative mechanism, given its prominence in the behavioral AI literature (Yam et al., 2023). Although mind perception may readily occur in any judgments (including of AI agents; Gray & Wegner, 2012), we predicted that it would not influence observers' attribution of responsibility in team settings because mind perception is merely an assessment of mental abilities (Gray et al., 2007), whereas perceiving autonomy centers on observers' evaluation of an entity's ability to act independently, a construct which encompasses team structural factors and their socio-normative evaluation (O'Neill et al., 2022; Wertenbroch et al., 2020). Though an artificial agent may be perceived to have significant mental capabilities, for a comprehensive assessment of moral responsibility observers may find it more important to judge whether it is "free" to do so (Alicke, 2000; Cushman, 2008; Malle et al., 2014). However, to rule out the alternative, Study 4 set out to test whether the AIHR effect holds even after controlling for mind perception.

*7.1. Method*

*7.1.1. Participants and design*

Six hundred and one US residents (US, UK, Ireland, Australia, Canada, and New Zealand nationalities, all English native speakers) were recruited from Prolific, an online research panel platform, to participate in the study. The participants ($M_{age}$=38.4 years; $SD_{age}$=13.1 years; 49% female, 50% male. 1% other) were randomly assigned to one of the four conditions in a 2 (team type: AI–Human vs. a Human–Human team) × 2 (perspective: Self vs. Other) between-subjects design.

*7.1.2. Experimental procedure*

Like in previous studies, participants responded to a scenario. In the self-perspective conditions, they were told "You are teamed up with…," and in the other-perspective conditions, "Jo is teamed up with…." Jo was used as a gender-neutral first name and positions the respondent as a third person observer. The team type factor manipulation was identical to that in previous studies, mentioning an "Artificial Intelligence system" or a colleague ("Employee 2") as a second team member. In all conditions, participants were told that the loan provider had received discriminatory lending complaints. Further, participants were told that an internal audit found that the team in focus was responsible for the mistake.

After reading the stimuli, participants attributed responsibility between involved parties using a slider, as before. For example, if participants had been told that they were teamed up with an AI system (self-perspective + AI condition), then participants attributed responsibility between themselves and the AI on a scale of 0 to 100 (counterbalanced in direction; three items; $\alpha = .87$). Participants then assessed the perceived mind of the counterparty in the team (AI or Employee 2) on four items ("capable of telling right from wrong," "understanding how others feel," "remembering things," "making plans"; $\alpha = .78$), using a seven-point Likert scale from 1 to 7 (i.e., 1= "not at all" and 7 = "very much"; adapted from Srinivasan & Sarial-Abi, 2021 and Sullivan & Fosso Wamba, 2022). At the end of the study, participants responded to demographic and two manipulation check questions. For team type, the categorical manipulation check question asked whether the other team member was an AI or a human. For perspective, the manipulation check asked who was assigned to review applications with the AI/colleague: "You" (first person perspective) or "Jo" (third person perspective).

*7.2. Results*

To test whether the team type manipulation worked, we regressed the team type manipulation check item on team type and perspective, including their interaction, using logistic regression. Team type strongly predicted perceived team type, $b = 6.36$, $SE = 0.61$, $z = 10.39$, $p < .001$, corresponding to an odds ratio of 579.17, 95% CI [194.94, 2236.17]. In contrast, perspective ($p = .498$) and the team type × perspective interaction ($p = .698$) were not significant, indicating that the perspective manipulation did not contaminate the team type manipulation check. Descriptively, participants' manipulation check responses were consistent with the assigned team type for 92.9% of those in the Human–Human condition and 97.4% of those in the AI–Human condition (misclassification rates 7.1% and 2.6%, respectively). For the perspective manipulation, we regressed the perspective manipulation-check item on perspective and team type, plus their interaction. Perspective was a very strong predictor of perceived perspective, $b = 6.22$, $SE = 0.58$, $z = 10.67$, $p < .001$, OR = 500.50, 95% CI [175.59, 1765.25], whereas team type ($p = .124$) and the perspective × team type interaction ($p = .177$) were not significant. Thus, the perspective manipulation check was unaffected by team type. In the Other Perspective condition, 96.4% of participants gave consistent responses consistent (3.6% inconsistent), and in the Self Perspective condition, 97.0% gave consistent responses (3.0% inconsistent). Overall, both experimental manipulations were successful and largely independent of each other.

To test the main effects and their possible interaction, we conducted a 2 × 2 ANOVA with team type and perspective as predictors and the average responsibility slider score as the outcome. Results showed significant main effects of team type, $F(1, 597) = 28.10$, $p < .001$, $p\eta^2 = .05$, and – consistent with actor-observer effects (Jones & Nisbett, 1971) – perspective, $F(1, 597) = 24.94$, $p < .001$, $p\eta^2 = .04$. The team type × perspective interaction was not significant, $F(1, 597) = 0.13$, $p = .716$, $p\eta^2 \approx .00$. Supporting our arguments and earlier findings, planned contrasts comparing team type levels within each perspective condition showed that participants attributed more responsibility to humans in the AI–Human than in the Human–Human team condition, from both the Other Perspective, $t(597) = 3.50$, $p < .001$, ($M_{AI–Human} = 64.10$, $SD_{AI–Human} = 23.31$ vs. $M_{Human–Human} = 55.08$, $SD_{Human–Human} = 16.88$; $d = .40$) and the Self Perspective, $t(597) = 4.00$, $p < .001$, ($M_{AI–Human} = 55.64$, $SD_{AI–Human} = 28.63$ vs. $M_{Human–Human} = 45.30$, $SD = 18.67$; $d = .46$).

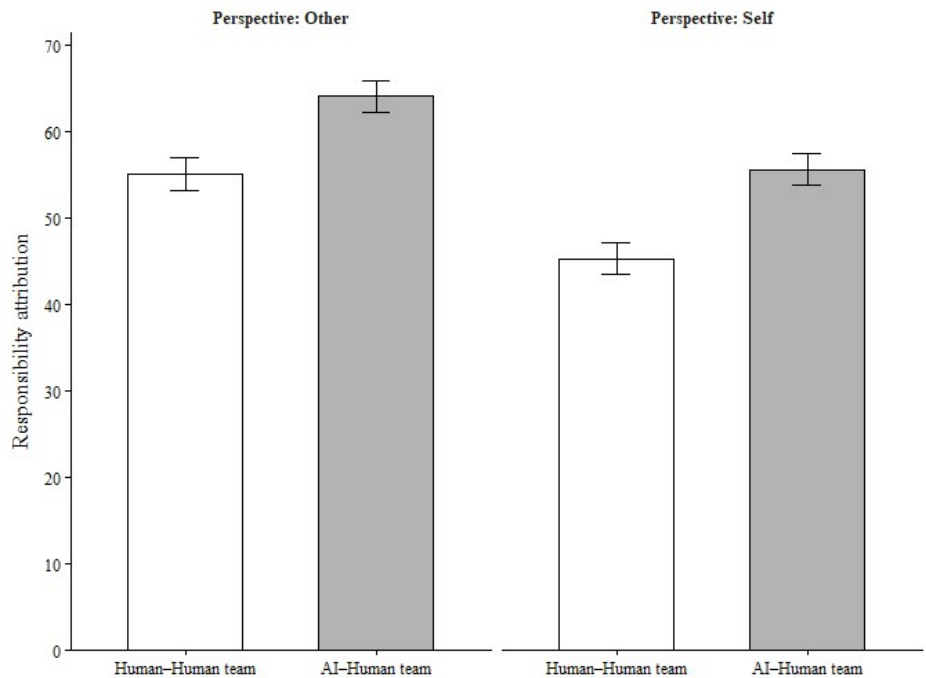

**Fig. 3.** Study 4: AI–Human team settings induce human responsibility, irrespective of the other–self difference in perspective.

Similar direct effects were observed in terms of respondents' perceiving mind in the teammate. To test the main effects and their possible interaction, we conducted a 2 × 2 ANOVA with team type and perspective as predictors and mind perception as the outcome, where higher values on the measure indicate more perceived mind in the teammate. Results showed a significant main effect of team type, $F(1, 597) = 149.53$, $p < .001$, $p\eta^2 = .20$, whereas the main effect of perspective was not significant, $F(1, 597) = 3.13$, $p = .078$, $p\eta^2 \approx .01$, and the team type × perspective interaction[1] was marginally significant and weak, $F(1, 597) = 3.61$, $p = .058$, $p\eta^2 \approx .01$. Planned contrasts comparing team type levels within each perspective condition showed that participants attributed less mind to the AI teammate in the AI–Human team condition than to the human teammate in the Human–

---

[1] The marginal interaction suggests that, though perceptions of AI mind did not change across perspective conditions, the human teammate was attributed somewhat more mind when participants imagined a team of two other people rather than their own partner. This finding is not important for our investigation and is not discussed further.

Human team condition, from both the Other Perspective, $t(597) = -10.0$, $p < .001$, ($M_{\text{AI–Human}} = 3.58$, $SD_{\text{AI–Human}} = 1.21$ vs. $M_{\text{Human–Human}} = 4.98$, $SD_{\text{Human–Human}} = 1.21$; $d = -1.15$) and the Self Perspective, $t(597) = -7.29$, $p < .001$, ($M_{\text{AI–Human}} = 3.59$, $SD_{\text{AI–Human}} = 1.18$ vs. $M_{\text{Human–Human}} = 4.62$, $SD_{\text{AI–Human}} = 1.28$; $d = -0.84$).

Crucially for our theoretical argument, mind perception did not mediate the focal AIHR effect. Mediation analysis using PROCESS Model 7 with 10,000 bootstrapping resamples (version 5.0, Hayes, 2025) estimated the first-stage model predicting mind perception from team type, perspective, and their interaction. Consistent with the ANOVA, team type significantly reduced teammate mind perception ($b = -1.78$, $SE = 0.31$, $p < .001$), and the team type × perspective interaction was weak and only marginal ($b = 0.38$, $SE = 0.20$, $p = .058$), with simple effects indicating lower mind perception in the AI–Human (vs. Human–Human) team condition under both the Other Perspective ($b = -1.41$, $SE = 0.14$, $p < .001$) and the Self Perspective ($b = -1.03$, $SE = 0.14$, $p < .001$). In the outcome model, mind perception did not significantly predict responsibility when controlling for team type ($b = 1.05$, $SE = 0.76$, $p = .168$), whereas the direct effect of team type on responsibility remained significant ($b = 11.00$, $SE = 2.08$, $p < .001$). Accordingly, the conditional indirect effects of team type on responsibility through mind perception were not significant (Other Perspective: $b = -1.48$, 95% BootCI [−4.01, 0.87]; Self Perspective: $b = -1.08$, 95% BootCI [−2.95, 0.64]). The index of moderated mediation was therefore not significant ($IMM = 0.40$, 95% BootCI [−0.25, 1.41]), indicating that mind perception did not mediate (nor differentially mediate across perspectives) the team type effect on responsibility.

*7.3. Discussion*

This last study replicated the effect of team type on the attribution of responsibility across two different perspectives (Self vs. Other). Whether participants imagined themselves or others teaming with AI, they attributed more responsibility to the human in the team. A closer look at the AI–Human team across the two perspectives shows that, even though self–other differences in attribution (Jones & Nisbett, 1971) are observable (people attribute less responsibility to themselves than others such as "Jo"), the AIHR effect holds robustly, trumping a well-established regularity in human behavior.

## 8. General discussion

Across four studies (total $N = 1,801$), we find converging evidence for AI-Induced Human Responsibility (AIHR). When a morally consequential mistake occurs under ambiguous joint responsibility, people allocate more responsibility to the human when the human is paired with AI than when paired with another human. The effect emerges in first-person/participant and third-person/observer perspectives and is not confined to severe moral harm; it holds for both high- and low-harm violations.

Taken together, these results suggest that AI–human teaming changes how responsibility is inferred (through perceptions of autonomy). Even when "blame the bot" heuristics and motivated self-exoneration should be readily available, participants shifted responsibility toward humans – including themselves – in AI–human teams. This occurred even though respondents were given no good reason to blame one teammate over the other – normatively, the response should have been a 50-50% random split. Indeed, they certainly should have had no reason to blame themselves, since the scenario made it clear they had not done anything wrong.

These findings are consistent with an attributional logic in which observers treat AI as comparatively constrained and, therefore, less plausible as a moral agent – making the human teammate the default locus of discretion and accountability, an intriguing positive "side-effect" of AI deployment in organizations.

*8.1. Theoretical Implications*

First, our findings introduce AIHR as a distinct responsibility-allocation pattern that complements (and complicates) dominant narratives in behavioral AI research that emphasize algorithm aversion (Benk et al., 2025) and "blame the bot" (Crolic et al., 2022). Rather than assuming that negative prior beliefs about AI straightforwardly translate into greater blame of AI after failures, AIHR shows that the hybrid AI-human teaming frame (joint action, partitioned accountability, and ambiguity about causal contribution; Schmutz et al., 2024; Choudhary et al., 2025) can flip intuitive predictions: AI can become the reason responsibility is pulled toward humans. This suggests that scholars should treat AI–human teaming as a qualitatively different social-cognitive context than human-AI-use-as-tool, with different default attribution rules.

Second, the strength and generality of AIHR suggest dominance over a powerful human tendency for self-serving bias or self-other differences in attributions (Darke & Chaiken, 2005; Larson, 1977; Jones & Nisbett, 1971). Across studies, participants reliably assigned more responsibility to the human in AI–human teams even when doing so is psychologically costly (e.g., first-person perspective) and despite well-established tendencies to protect the self. This implies that motivational accounts of blame shifting in organizations may be incomplete without considering how the composition of the action system (human–human vs. AI–human) changes what "deflection" looks like. Put differently, AI may create a responsibility landscape in which evasion is less plausible because the human is perceived as the *only* party with legitimate moral discretion, increasing human responsibility.

Third, our evidence clarifies the psychological mechanism of moral attribution in AI-human teams by separating perceived autonomy (Bacharach & Aiken, 1976; Bigman et al., 2019) from mind perception (Yam et al., 2023). Perceived autonomy mediates AIHR, whereas mind perception – despite differing strongly by condition – does not mediate responsibility allocation. This distinction matters theoretically because it suggests that responsibility judgments in organizational AI–human teaming contexts rely less on whether the AI agent seems "mindful" or cognitively capable and more on whether it is seen as structurally free to choose. This invites OB theory to conceptualize autonomy in AI-human systems as socially interpreted inferences about structural constraints (what an AI has perceived authority to do), not merely as a cognitive assessment of a technology feature (what it is perceived to be mentally capable of doing).

*8.2. Practical Implications*

For managers deploying AI into morally sensitive decisions (e.g., lending, hiring, compliance), AIHR implies that human employees may shoulder disproportionate perceived accountability when outcomes are contested or harmful. In practice, this can be beneficial if it prevents "responsibility gaps," but it also risks unfair blame, moral injury, and distorted learning from failures – especially when errors are meaningfully driven by system-level or model-level factors. Organizations should anticipate that simply pairing an employee with AI can implicitly reposition that employee as the primary bearer of responsibility.

Second, AIHR underscores the importance of explicit role design and decision-right clarity in AI-human teams. If perceived autonomy drives responsibility, then governance choices (who has override authority, how escalation works, what is mandatory vs. advisory, and how model recommendations are logged) are not mere technicalities. Since the mere pairing of humans with AI ascribe counterparty responsibility to humans (even when unwarranted), explicit design needs to complement (or override) this newly discovered moral heuristic. Clear documentation of decision pathways and structured review procedures can help align perceived responsibility with actual control and prevent over-internalization by frontline employees.

Third, AIHR suggests implications for organizational culture, training, and ethics infrastructure. Organizations often fear moral disengagement and "hands off the wheel" outcomes when it comes to technology deployment and they encourage employees to "own the outcome" when using AI (Ogunfowora et al., 2022; Parasuraman & Manzey, 2010). Our results suggest that this message may already be psychologically amplified. Training should therefore include not only "how to use AI" but also "how to attribute responsibility appropriately," including when to question, override, or escalate AI-supported decisions. Designing psychologically safe reporting channels for AI-related near misses – and framing responsibility as shared and diagnostic rather than purely punitive – may help prevent AIHR from becoming a pathway to scapegoating or burnout.

## 8.3. Limitations and future directions

A first limitation is that our evidence comes from scenario-based experiments in a single institutional domain (banking/loan processing). Though this choice offers control and clear moral stakes, it leaves open questions about generalizability to other tasks (e.g., healthcare, policing, creative work) and to contexts where AI systems are more visibly agentic (e.g., conversational LLM copilots that appear initiative-taking or emerging "agentic AI" tools; Li et al., in press). Future work should replicate AIHR across industries and with richer task ecologies, including settings where outcomes unfold over time and responsibility is contested in real organizational processes.

Second, our dependent measures capture stated responsibility attributions rather than downstream behaviors (e.g., willingness to report errors, accept sanctions, apologize, compensate stakeholders, or disengage morally). This matters because responsibility judgments can be performative and context-dependent. Research that links AIHR to behavioral outcomes (such as escalation, blame communication, learning behaviors, or ethical fading) would clarify whether AIHR changes behaviors beyond stated assessments.

Third, future research should map boundary conditions tied to autonomy cues. If perceived autonomy is pivotal, then AIHR should vary systematically with design and context features that signal discretion: explainability, explicit "approval required" policies, the salience of human override, the AI's degree of initiative, or organizational messaging that frames AI as advisor vs. co-decider. Such work could also test when AIHR reverses (e.g., when AI is framed as having broad delegated authority or when humans are described as procedurally constrained).

Finally, the broader theoretical agenda is to study AIHR as part of a dynamic, multilevel, hybrid accountability system. Real organizations involve hierarchies, cross-functional teams, vendors, and regulators – each of which can shape where responsibility "lands." Future studies could examine triads or larger teams (e.g., human–human–AI or even much larger networks, using network analysis), investigate how responsibility shifts across hierarchical levels, and test how organizational norms (e.g., "AI use is mandatory," "AI decisions are audited," "AI is organization-approved") amplify or attenuate AIHR over time. This would help build more complete theories of how organizations distribute moral responsibility as AI becomes a routine teammate rather than an executional tool.

# WEB APPENDIX

# AI-Induced Human Responsibility (AIHR) in AI–human teams

## Web Appendix A: Survey Design, Stimuli, and Measures for Study 1

Study 1 used a single variable with two levels (team type: AI–Human vs. Human–Human) between-subjects design. Participants, recruited from Prolific, an online sample provider, were randomly assigned to one of the two experimental conditions. The study was administered online using Qualtrics, an online survey platform.

*Stimuli*

Participants started by reading one of the stimuli, as seen below.

| Team type: AI–Human | Team type: Human–Human |
|---|---|
| Imagine you work for Bank X, in its credit and loan department.<br><br>**You are teamed up with an Artificial Intelligence system** and responsible for assessing loan applications and then whether to approve them.<br><br>**For each loan application, the team (i.e., you and the Artificial Intelligence system) is responsible for evaluating the requested loan amount against relevant documents, such as identity documents as well as payslips, and deciding whether to accept or change the proposed loan amount.**<br><br>Then, Bank X provides the loan amount to clients based on the team's decision. | Imagine you work for Bank X, in its credit and loan department.<br><br>**You are teamed up with a colleague** and responsible for assessing loan applications and then whether to approve them.<br><br>**For each loan application, the team (i.e., you and your colleague) is responsible for evaluating the requested loan amount against relevant documents, such as identity documents as well as payslips, and deciding whether to accept or change the proposed loan amount.**<br><br>Then, Bank X provides the loan amount to clients based on the team's decision. |

Then, after proceeding to the next page, participants in both conditions read an additional, identical piece of information, as seen below.

> Later you learn that Bank X received discriminatory lending complaints.
>
> **Loan clients were unfairly treated as their borrowing limit was based on their backgrounds rather than loan serviceability.**
>
> An internal audit determined **the team that assessed loan applications was responsible for the mistake**, as deciding clients' borrowing amounts is within the team's scope of work.

*Measurement*

After reading the stimulus, participants answered a series of questions (see below). For the key DV, attribution of responsibility we used 0-100 sliders, counterbalancing the anchors to rule out order effects (Lensvelt-Mulders et al., 2005):

- 0 = "You" to 100 = "The Artificial Intelligence system/Colleague" (RS slider)

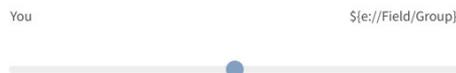

- Or 0 = "The Artificial Intelligence system/Colleague" to 100 = "You" (LS slider)

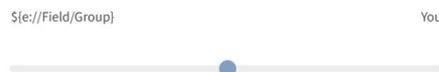

| Construct | Scale | Operationalization |
|---|---|---|
| *Dependent variable* | | |
| Responsibility attribution (adapted from Hinds et al., 2004; Lei & Rau, 2021) | Counter-balanced slider from 0 to 100 | • Thinking about your team, using the slider, indicate the party that you think would be more responsible for the outcome<br>(See counterbalancing above; half of the responses [RS slider] were transformed [=100 – raw value] such that a higher value indicates more responsibility attributed to the Self.) |
| *Control variables* | | |
| Perception of team performance (Aubé & Rousseau, 2005) | 7-point Likert scales | • I am confident in the team's (i.e., you and the Artificial Intelligence system/colleague) capabilities;<br>("1 = disagree strongly, and 7 = agree strongly")<br>• I respect the team (i.e., you and the Artificial Intelligence system/colleague);<br>("1 = disagree strongly, and 7 = agree strongly")<br>• To what extent, do you rate the team's (i.e., you and the Artificial Intelligence/colleague) capabilities in completing loan limit assessment tasks going forward?;<br>("1 = not at all capable, and 7 = very capable") |
| Positive self-image (Marlowe Social Desirability scale (Ventimiglia & MacDonald, 2012) | 7-point Likert scales | • I always tell the truth, no matter what;<br>("1 = disagree strongly, and 7 = agree strongly")<br>• I always try to be fair in my dealings with others;<br>("1 = disagree strongly, and 7 = agree strongly")<br>• I never get angry or lose my temper;<br>("1 = disagree strongly, and 7 = agree strongly")<br>• I never cover up my mistakes;<br>("1 = disagree strongly, and 7 = agree strongly") |

*Demographic and manipulation check items*

Shortly after, participants responded to general and demographic-related questions.

| Item | Operationalization |
|---|---|
| *Demographic* | |
| Gender | • What is your gender? (Male, Female, Prefer to self-identify) |
| Age | • What is your age |
| Political philosophy | • How would you classify yourself in terms of political affiliation?; ("Liberal", "Conservative", "Other") |
| *Manipulation check item* | |
| Team type | • Thinking back on the scenario you read, who else did Bank X assign to review loan applications with you?<br>  ○ An Artificial Intelligence (AI)<br>  ○ A human |

*Additional analyses*

We conducted further robustness tests. First, for the main hypothesis, Levene's test indicated that the assumption of homogeneity of variance was violated, with significantly different DV variances across conditions, $F(1, 300) = 26.31$, $p < .001$. Consequently, we repeated the main hypothesis test, using Welch's t, which yielded virtually identical results, $t_{\text{Welch}}(274.57) = 2.12$, $p = .035$.

Next, we also repeated the covariate tests (using perceived team performance and social desirability as additional predictors) with heteroscedasticity-robust (HC3) standard errors. Regressing responsibility on team type (dummy coded), perceived team performance (mean-centered), and their interaction again revealed a significant effect of team performance, indicating that higher team performance was associated with higher human responsibility scores, $b = 4.64$, $SE_{HC3} = 1.43$, $t = 3.25$, $p = .001$. Importantly, the main effect of team type remained significant, such that participants in the AI–Human condition reported higher human responsibility scores, as before $b = 7.75$, $SE_{HC3} = 3.20$, $t = 2.43$, $p = .016$, and the interaction was not significant, $b = 2.24$, $SE_{HC3} = 2.40$, $t = .93$, $p = .352$.

Then, using heteroscedasticity-robust (HC3) standard errors again, we regressed responsibility on team type (dummy coded), social desirability (mean-centered), and their interaction. We found a non-significant effect of social desirability, $b = 1.84$, $SE_{HC3} = 2.11$, $t = .87$, $p = .386$. Importantly, the main effect of team type remained significant, such that participants in the AI–Human condition reported higher human responsibility scores than in the Human–Human condition, $b = 7.01$, $SE_{HC3} = 3.33$, $t = 2.10$, $p = .036$), and the interaction was not significant, $b = –4.01$, $SE_{HC3} = 3.54$, $t = –1.13$, $p = .258$. This provides further evidence that social desirability does not interfere with our results.

We examined whether demographic variables (age, gender, political philosophy) may confound the association between team type and attribution of responsibility. Zero-order correlations among *age* showed non-significant correlations with either team type, $r(302) = .09$, $p = .126$, or responsibility, $r(302) = .02$, $p = .761$. Entering it as a moderator in our core model did not yield a main, $b = –.13$, $SE_{HC3} = .15$, $t = –.87$, $p = .386$, or interactive effect, $b = .26$, $SE_{HC3} = .23$, $t = 1.14$, $p = .255$, leaving our focal prediction intact, $b = 7.01$, $SE_{HC3} = 3.36$, $t = 2.09$, $p = .038$.

To examine whether *gender* might moderate or account for the effect of team type on the DV, we re-ran the analysis after excluding participants who selected the third gender option (leaving N = 299). Levene's test indicated heterogeneity of variances across the IV × gender cells, $F(3, 295) = 11.78$, $p < .001$, so we relied on HC3-robust standard errors. A 2 (AI–Human vs Human–Human team) × 2 (gender) model including the interaction showed no evidence that the effect of team type depended on gender, $F_{robust}(1, 295) = .00$, $p = .947$. We therefore fit an additive model with team type and gender as predictors. In this robust model, the effect of team type on DV remained significant, $F_{\text{robust}}(1, 296) = 4.64$, $p = .032$, with participants in the AI–Human condition reporting high DV scores than those in the Human–Human condition, $b = 7.26$, $SE_{HC3} = 3.37$, $t(296) = 2.16$, $p = .032$. By contrast, gender was not a significant predictor of DV, $F_{\text{robust}}(1, 296) = 0.39$, $p = .535$, $b = -2.10$, $SE_{HC3} = 3.38$, $t(296) = -0.62$, $p = .535$. Thus, gender neither moderates nor meaningfully explains the effect of team type on attribution of human responsibility.

We also tested whether political philosophy (PolPhi; Liberal, Conservative, Other) should be included alongside team type when predicting the DV. Levene's test again indicated unequal variances across the IV × PolPhi cells, $F_{\text{robust}}(5, 296) = 5.71$, $p < .001$, so we used HC3-robust standard errors. In the full model, the IV × PolPhi interaction was not significant, $F_{\text{robust}}(2, 296) = 0.01$, $p = .992$, indicating that the effect of IV did not differ by political philosophy. In the additive model (Type II), IV remained a significant predictor of DV, $F_{\text{robust}}(1, 298) = 3.99$, $p = .047$, corresponding to an adjusted difference of 6.64 points between conditions ($SE_{HC3} = 3.32$), $t(298) = 2.00$, $p = .047$. Political philosophy was not a reliable predictor overall, $F_{\text{robust}}(2, 298) = 2.38$, $p = .094$. Using Liberal as the reference group, neither Conservative, $b = 6.96$, $SE_{HC3} = 4.12$, $t(298) = 1.69$, $p = .092$, nor Other, $b = -3.34$, $SE_{HC3} = 4.16$, $t(298) = -0.80$, $p = .423$ differed significantly from Liberals. Holm-adjusted pairwise comparisons likewise showed no significant differences among the three PolPhi groups (all $ps \geq .108$). Overall, political philosophy provided at most weak evidence of association with DV and did not change our focal effect. Since similar analyses for demographic variables did not have confounding effects in Studies 2, 3 and 4 either, we will not discuss these further.

Finally, we wanted to test that there were order effects in the sliders, using our counterbalancing factor. Levene's test showed heteroskedasticity, $F(3, 298) = 9.27, p < .001$. We therefore ran a heteroscedasticity-robust 2 (team type: IV) × 2 (slider order: "You" on the right vs left) ANOVA on responsibility ratings. The main effect of counterbalancing was not significant, $F_{\text{robust}}(1, 298) = .13, p = .718$, and the IV × counterbalancing interaction was likewise non-significant, $F_{\text{robust}}(1, 298) = 0.25, p = .614$, while the effect of team type was marginal, $F_{\text{robust}}(1, 298) = 3.63, p = .058$ (noting that entering nonsignificant independent terms and their interactions can unduly reduce significance of true predictors). Taken together, these results indicate that neither the overall level of responsibility nor the AIHR effect itself depended on the order in which the sliders were presented.

**Web Appendix B: Survey Design, Stimuli, and Measures for Study 2**

Study 2 was conducted to test the perception of autonomy as the mechanism behind the effect of team arrangement on the attribution of responsibility with a 2 (team type: AI–Human team vs. Human–Human team) × 2 (harm level: High vs. Low) between-subjects design. Participants recruited from Prolific, were randomly assigned to one of the four experimental conditions. The study was administered online using Qualtrics, an online survey platform.

*Stimuli*

Depending on their assigned group, participants read one of the four stimuli in this study as seen below.

| | Team type: AI–Human | Team type: Human–Human |
|---|---|---|
| Harm level: High | Employee 1 works for Bank X's credit and loan department. **Employee 1 is teamed up with an Artificial Intelligence system** and responsible for assessing loan applications and then whether to approve them.<br><br>**For each loan application, the team (i.e., Employee 1 with the Artificial Intelligence system) is responsible for evaluating the requested loan amount against relevant documents, such as identity documents as well as payslips, and deciding whether to accept or change the proposed loan amount.**<br>Then, Bank X provides the loan amount to clients based on the team's decision.<br><br>========PAGE BREAK=======<br><br>Later, you learn that Bank X's internal compliance department found an instance of financial regulation violation in these loan approvals.<br>An audit determined **the team that assessed loan applications was responsible for the error**, as deciding clients' borrowing amounts is within the team's scope of work.<br>The audit found that **the team's falsification of the client's financial information led to the approval and provision of loan amounts that were beyond the client's loan serviceability, causing financial hardship.** | Employee 1 works for Bank X's credit and loan department. **Employee 1 is teamed up with Employee 2** and responsible for assessing loan applications and then whether to approve them.<br><br>**For each loan application, the team (i.e., Employee 1 with Employee 2) is responsible for evaluating the requested loan amount against relevant documents, such as identity documents as well as payslips, and deciding whether to accept or change the proposed loan amount.**<br>Then, Bank X provides the loan amount to clients based on the team's decision.<br><br>========PAGE BREAK=======<br><br>Later, you learn that Bank X's internal compliance department found an instance of financial regulation violation in these loan approvals.<br>An audit determined **the team that assessed loan applications was responsible for the error**, as deciding clients' borrowing amounts is within the team's scope of work.<br>The audit found that **the team's falsification of the client's financial information led to the approval and provision of loan amounts that were beyond the client's loan serviceability, causing financial hardship.** |

|  | **Team type: AI–Human** | **Team type: Human–Human** |
|---|---|---|
| Harm level: Low | Employee 1 works for Bank X's credit and loan department. **Employee 1 is teamed up with an Artificial Intelligence system** and responsible for assessing loan applications and then whether to approve them. **For each loan application, the team (i.e., Employee 1 with the Artificial Intelligence system) is responsible for evaluating the requested loan amount against relevant documents, such as identity documents as well as payslips, and deciding whether to accept or change the proposed loan amount.** Then, Bank X provides the loan amount to clients based on the team's decision.<br><br>========PAGE BREAK=======<br><br>Later, you learn that Bank X's records management department found instances of filing errors in these loan approvals. An audit determined **the team that assessed loan applications was responsible for the error**, as filing clients' forms into the bank's system is within the team's scope of work. The audit found that **the team's mistakes in assigning file numbers on loan applications for some clients led to additional work for another team to rectify the clients' file numbers.** | Employee 1 works for Bank X's credit and loan department. **Employee 1 is teamed up with Employee 2** and responsible for assessing loan applications and then whether to approve them.<br><br>**For each loan application, the team (i.e., Employee 1 with Employee 2) is responsible for evaluating the requested loan amount against relevant documents, such as identity documents as well as payslips, and deciding whether to accept or change the proposed loan amount.** Then, Bank X provides the loan amount to clients based on the team's decision.<br><br>========PAGE BREAK=======<br><br>Later, you learn that Bank X's records management department found instances of filing errors in these loan approvals. An audit determined **the team that assessed loan applications was responsible for the error**, as filing clients' forms into the bank's system is within the team's scope of work. The audit found that **the team's mistakes in assigning file numbers on loan applications for some clients led to additional work for another team to rectify the clients' file numbers.** |

*Measurement*

After reading the stimulus, participants answered a series of questions below. For question items using a slider, we randomized the appearance order of the label on the slider to rule out order effects.

| Construct | Scale | Operationalization |
|---|---|---|
| *Dependent variable* <br> Responsibility attribution (adapted from Hinds et al., 2004; Lei & Rau, 2021) | A slider with value 0 to 100 | • Thinking about the team, using the slider, indicate whom the outcome was primarily dependent on.; ("0 = Employee1, and 100 = the Artificial Intelligence/ Employee 2") <br> • Thinking about the team, using the slider, indicate who was the most responsible for the outcome.; ("0 = Employee1, and 100 = the Artificial Intelligence/ Employee 2") <br> • Thinking about the team, using the slider, indicate who should assume most of the punishment for the outcome.; ("0 = Employee1, and 100 = the Artificial Intelligence/ Employee 2") |
| *Mediator* <br> Perception of autonomy (adapted from Bacharach & Aiken, 1976; Berretta et al., 2023) | A slider with value 0 to 100 | • When assessing the loan applications, which team member could have had more amount of freedom in terms of planning and performing the tasks?; ("0 = Employee1, 50 = equal, and 100 = the Artificial Intelligence/ Employee 2") <br> • When assessing the loan applications, which team member could have had more opportunity to make decisions without consulting anyone else?; ("0 = Employee1, 50 = equal, and 100 = the Artificial Intelligence/ Employee 2") <br> • When assessing the loan applications, which team member could have had more independence in deciding things?; ("0 = Employee1, 50 = equal, and 100 = the Artificial Intelligence/ Employee 2") <br> • When assessing the loan applications, which team member could have had more say in how to |

| | |
|---|---|
| | manage the tasks?; ("0 = Employee1, 50 = equal, and 100 = the Artificial Intelligence/ Employee 2")
• When assessing the loan applications, which team member could have had more autonomy to do as they please?; ("0 = Employee1, 50 = equal, and 100 = the Artificial Intelligence/ Employee 2")
• When assessing the loan applications, which team member could have had more choice in determining how things should have been done?; ("0 = Employee1, 50 = equal, and 100 = the Artificial Intelligence/ Employee 2") |

***Demographic and manipulation check items***

Shortly after, participants responded to general and demographic-related questions, as seen below.

| Item | Operationalization |
|---|---|
| *Demographic* | |
| Gender | • What is your gender? (Male, Female, Prefer to self-identify) |
| Age | • What is your age? |
| | |
| *Manipulation check item* | |
| Team type | • Thinking back on the scenario you read, who else did Bank X assign to review loan applications with Employee 1?<br>  ○ An Artificial Intelligence (AI) agent<br>  ○ Employee 2 |
| Harm level | • Thinking back on the scenario you read, what was the audit's allegation on the team?<br>  ○ Falsification of clients' financial information<br>  ○ Filing error |

***Additional analyses***

Because a Levene's test indicated substantial heterogeneity of variance across the four experimental cells with regards to the main dependent variable, $F(3, 598) = 52.20$, $p < .001$, we re-estimated the 2 × 2 ANOVA using a model with HC3 heteroskedasticity-robust standard errors (White correction, Type III). The pattern of results was unchanged. The main effect of team type remained significant, $F_{robust}(1, 598) = 65.15$, $p < .001$, $p\eta^2 = .10$, whereas the main effect of harm level was not significant, $F_{robust}(1, 598) = 0.85$, $p = .356$, $p\eta^2 \approx .00$, and the interaction between team type and harm was also nonsignificant, $F_{robust}(1, 598) = 0.47$, $p = .500$, $p\eta^2 \approx .00$. Similarly, a Levene's test indicated heterogeneity of variance with regards to the mediator, perceived autonomy, as well, $F(3, 598) = 48.83$, $p < .001$. A robust ANOVA showed that the main effect of team type was significant, $F_{robust}(1, 598) = 188.24$, $p < .001$, $p\eta^2 = .24$, whereas the main effect of harm level was not significant, $F_{robust}(1, 598) = .08$, $p = .778$, $p\eta^2 \approx .00$, and the interaction between team type and harm was also nonsignificant, $F_{robust}(1, 598) = 0.86$, $p = .353$, $p\eta^2 \approx .00$.

We also tested the possible effects of counterbalancing. A Levene's test including team type, harm, and counterbalancing indicated substantial heterogeneity of variance across the eight factorial cells, $F(7, 594) = 21.61$, $p < .001$. We therefore estimated a 2 (team type: AI–Human vs Human–Human) × 2 (harm: High vs Low) × 2 (counterbalancing: Employee 1 on the left vs right) ANOVA on responsibility using HC3 heteroskedasticity-robust (White-corrected, Type III) standard errors. The robust ANOVA replicated the main effect of team type, $F_{robust}(1, 594) = 65.68$, $p < .001$, $p\eta^2 = .10$, while the main effect of counterbalancing was not significant, $F_{robust}(1, 594) = 0.23$, $p = .63$, $p\eta^2 \approx .00$. Moreover, counterbalancing did not interact with team type or harm, nor was the three-way interaction significant, all $F_{robust}(1, 594) \leq 1.49$, all $p \geq .23$, $p\eta^2 \approx .00$. Similar findings emerged for treating perceived autonomy as a dependent variable in that the robust ANOVA showed a strong main effect of team type, $F_{robust}(1, 594) = 198.53$, $p < .001$, $p\eta^2 = .25$, whereas neither harm nor counterbalancing had significant main effects, $F_{robust}(1, 594) = 0.44$, $p = .51$, $p\eta^2 \approx .00$, and $F_{robust}(1, 594) = 2.00$, $p = .16$, $p\eta^2 \approx .00$, respectively. However, team type interacted with counterbalancing, $F_{robust}(1, 594) = 9.07$, $p$

= .003, $p\eta^2$ = .01, while the remaining interactions (including the three-way) were not significant, all $F_{robust}(1, 594) \leq 3.38$, all $p \geq .066$, $p\eta^2 \approx .01$.

Including both counterbalancing and harm level as covariates in the PROCESS Model 4 analysis did not meaningfully change the mediation pattern. Counterbalancing was not a significant predictor of perceived autonomy ($b = 1.77$, $p = .175$) or responsibility (total model: $b = -0.70$, $p = .647$; responsibility model: $b = -1.24$, $p = .402$). Likewise, harm type was not a significant predictor of perceived autonomy ($b = -0.53$, $p = .682$) or responsibility (total model: $b = -1.31$, $p = .388$; responsibility model: $b = -1.15$, $p = .434$). The indirect effect of team type on responsibility via perceived autonomy remained significant ($b = 5.40$, $BootSE = 1.24$, 95% CI [3.13, 7.95]). Thus, neither the ordering of the sliders nor the harm level manipulation meaningfully influenced the mediation effect.

**Web Appendix C: Survey Design, Stimuli, and Measures for Study 3**

Study 3 used a single variable with two levels (team type: AI–Human vs. Human–Human team) between-subjects design. Participants, recruited from Prolific, were randomly assigned to one of the two experimental conditions. The study was administered online using Qualtrics, an online survey platform.

*Stimuli*

Participants started by reading one of the stimuli, as seen below.

| Team: AI–Human | Agent: Human–Human |
|---|---|
| Imagine you work for Bank X's credit and loan department.<br><br>**You are teamed up with an Artificial Intelligence system** and responsible for assessing loan applications and then whether to approve them.<br><br>**For each loan application, the team (i.e., You with the Artificial Intelligence system) is responsible for evaluating the requested loan amount against relevant documents, such as identity documents as well as payslips, and deciding whether to accept or change the proposed loan amount.**<br><br>Then, Bank X provides the loan amount to clients based on the team's decision. | Imagine you work for Bank X's credit and loan department.<br><br>**You are teamed up with Employee 1** and responsible for assessing loan applications and then whether to approve them.<br><br>**For each loan application, the team (i.e., You with Employee 1) is responsible for evaluating the requested loan amount against relevant documents, such as identity documents as well as payslips, and deciding whether to accept or change the proposed loan amount.**<br><br>Then, Bank X provides the loan amount to clients based on the team's decision. |

Then, after proceeding to the next page, participants in both conditions read an additional piece of information, as seen below.

> Later you learn that Bank X's internal compliance department found an instance of financial regulation violation in these loan approvals.
>
> An audit determined **the team that assessed loan applications was responsible for the error**, as deciding clients' borrowing amounts is within the team's scope of work.
>
> The audit found that **the team's mistakes in evaluating the client's financial information led to the approval and provision of loan amounts that were beyond the client's loan serviceability, causing financial hardship.**

*Measurement*

After reading the stimulus, participants answered a series of questions below. For question items using a slider, we randomized the appearance order of the label on the slider to rule out order effects.

| Construct | Scale | Operationalization |
| --- | --- | --- |
| *Dependent variable* Responsibility attribution (adapted from Hinds et al., 2004; Lei & Rau, 2021) | A slider with value 0 to 100 | • Thinking about the team, using the slider, indicate whom the outcome was primarily dependent on.; ("0 = You 1, 50 = equal, and 100 = the Artificial Intelligence/ Employee 1") <br> • Thinking about the team, using the slider, indicate who was the most responsible for the outcome.; ("0 = You, 50 = equal, and 100 = the Artificial Intelligence/ Employee 1") <br> • Thinking about the team, using the slider, indicate who should assume most of the punishment for the outcome.; ("0 = You, 50 = equal, and 100 = the Artificial Intelligence/ Employee 1") |
| *Alternative explanation* | | |
| Threat to self-worth (adapted from Granulo et al., 2019) | A Likert scale with value 1 to 7 | • Working with [the Artificial intelligence/ Employee 1] makes me feel devalued.; ("1 = strongly disagree, 7 = strongly agree") <br> • Working with [the Artificial intelligence/ Employee 1] raises doubts about myself.; ("1 = strongly disagree, 7 = strongly agree") <br> • Working with [the Artificial intelligence/ Employee 1] makes me question my abilities.; ("1 = strongly disagree, 7 = strongly agree") |

### *Demographic and manipulation-check items*

Shortly after, participants responded to general and demographic-related questions, as seen below.

| Item | Operationalization |
|---|---|
| *Demographic* | |
| Gender | • What is your gender? (Male, Female, Prefer to self-identify) |
| Age | • What is your age |
| | |
| *Manipulation check item* | |
| Team type | • Thinking back on the scenario you read, who else did Bank X assign to review loan applications with you? <br>     o An Artificial Intelligence (AI) agent <br>     o Employee 1 |

### *Additional analyses*

Testing the main hypothesis, we again conducted a check on homoskedasticity. Levene's test indicated substantial heterogeneity of variance across the two experimental cells with regards to the main dependent variable, $F(1, 294) = 52.98$, $p < .001$. Consequently, we repeated the main hypothesis test, using Welch's t, which yielded identical results, $t_{\text{Welch}}(252.52) = 4.50$, $p < .001$.

The possible mediator, perceived threat to self-worth was homoscedastic with regard to the manipulation, $F(1, 294) = 1.29$, $p = .257$.

We also tested possible order effects in the sliders, using our counterbalancing factor. Levene's test showed heteroskedasticity, $F(3, 292) = 18.41$, $p < .001$. We therefore ran a heteroscedasticity-robust 2 (team type: IV) × 2 (slider order: "You" on the right vs left) ANOVA on responsibility ratings. The main effect of counterbalancing was not significant, $F_{robust}(1, 292) = .38$, $p = .537$, and the IV × counterbalancing interaction was likewise non-significant, $F_{robust}(1, 292) = .77$, $p = .381$, while the effect of team type was, $F_{robust}(1, 292) = 6.82$, $p = .009$. This indicates that neither the overall level of responsibility nor the AIHR effect itself depended on the order in which the sliders were presented.

**Web Appendix D: Survey Design, Stimuli, and Measures for Study 4**

Study 4 was conducted with a 2 (team: AI–Human vs. Human–Human) x 2 (Perspective: self vs. other) between-subjects design. Participants, recruited from Prolific, were randomly assigned to one of the four experimental conditions. The study was administered online using Qualtrics, an online survey platform.

*Stimuli*

Depending on their assigned group, participants read one of the four stimuli in this study as seen below.

|  | **Team: AI–Human** | **Team: Human–Human** |
| --- | --- | --- |
| Perspective: self | Imagine you work for Bank X, in its credit and loan department. **You are teamed up with an Artificial Intelligence system** and responsible for assessing loan applications and then whether to approve them. **For each loan application, the team (i.e., you and the Artificial Intelligence system) is responsible for evaluating the requested loan amount against relevant documents, such as identity documents as well as payslips, and deciding whether to accept or change the proposed loan amount.** Then, Bank X provides the loan amount to clients based on the team's decision. | Imagine you work for Bank X, in its credit and loan department. **You are teamed up with a colleague** and responsible for assessing loan applications and then whether to approve them. **For each loan application, the team (i.e., you and your colleague) is responsible for evaluating the requested loan amount against relevant documents, such as identity documents as well as payslips, and deciding whether to accept or change the proposed loan amount.** Then, Bank X provides the loan amount to clients based on the team's decision. |

| Perspective: other | Jo works for Bank X, in its credit and loan department. **Jo is teamed up with an Artificial Intelligence system** and responsible for assessing loan applications and then whether to approve them. **For each loan application, the team (i.e., Jo and the Artificial Intelligence system) is responsible for evaluating the requested loan amount against relevant documents, such as identity documents as well as payslips, and deciding whether to accept or change the proposed loan amount.** Then, Bank X provides the loan amount to clients based on the team's decision. | Jo works for Bank X, in its credit and loan department. **Jo is teamed up with a colleague** and responsible for assessing loan applications and then whether to approve them. **For each loan application, the team (i.e., Jo and their colleague) is responsible for evaluating the requested loan amount against relevant documents, such as identity documents as well as payslips, and deciding whether to accept or change the proposed loan amount.** Then, Bank X provides the loan amount to clients based on the team's decision. |

Then, after proceeding to the next page, participants in both conditions read an additional piece of information, as seen below.

| |
|---|
| Later you learn that Bank X received discriminatory lending complaints. **Loan clients were unfairly treated as their borrowing limit was based on their backgrounds rather than loan serviceability.** An internal audit determined **the team that assessed loan applications was responsible for the mistake**, as deciding clients' borrowing amounts is within the team's scope of work. |

*Measurement*

| Construct | Scale | Operationalization |
|---|---|---|
| *Dependent variable* | | |
| Responsibility attribution (adapted from Hinds et al., 2004; Lei & Rau, 2021) | A slider with value 0 to 100 | • Thinking about the team, using the slider, indicate whom the outcome was primarily dependent on.; ("0 = You/ Jo, and 100 = the Artificial Intelligence/ colleague") <br> • Thinking about the team, using the slider, indicate who was the most responsible for the outcome.; ("0 = You/ Jo, and 100 = the Artificial Intelligence/ colleague") <br> • Thinking about the team, using the slider, indicate who should assume most of the punishment for the outcome.; ("0 = You/ Jo, and 100 = the Artificial Intelligence/ colleague") |
| *Alternative explanation* | | |
| Mind perception (adapted from Srinivasan & Sarial-Abi, 2021 and Sullivan & Fosso Wamba, 2022) | A Likert scale with values from 1 to 7 | • To what extent do you think [the Artificial Intelligence/ colleague] is capable of telling right from wrong?; ("1 = not at all, 7 = very much") <br> • To what extent do you think [the Artificial Intelligence/ colleague] is capable of understanding how others feel?; ("1 = not at all, 7 = very much") <br> • To what extent do you think [the Artificial Intelligence/ colleague] is capable of remembering things?; ("1 = not at all, 7 = very much") <br> • To what extent do you think [the Artificial Intelligence/ colleague] is capable of making plans?; ("1 = not at all, 7 = very much") |

*Demographic and manipulation-check items*

Shortly after, participants responded to general and demographic-related questions, as seen below.

| Item | Operationalization |
| --- | --- |
| *Demographic* | |
| Gender | • What is your gender? (Male, Female, Prefer to self-identify) |
| Age | • What is your age |
| | |
| *Manipulation check item* | |
| Team type | • Thinking back on the scenario you read, who else did Bank X assign to review loan applications with you/Jo?<br>   o An Artificial Intelligence (AI) agent<br>   o A Human |
| Perspective | • Thinking back on the scenario you read, who did Bank X assign to review loan applications along with the Artificial Intelligence system/the colleague?<br>   o You<br>   o A person named Jo |

*Additional analyses*

As before, we checked heteroskedasticity. In terms of the main test variable, DV, Levene's test indicated heterogeneity of variances across the four cells (IV × Perspective), $F(3, 597) = 19.38$, $p < .001$. Consequently, we ran a heteroscedasticity-robust 2 (team type: IV) × 2 (perspective: self vs. other) ANOVA on responsibility ratings. The effect of team type, $F_{robust}(1, 597) = 28.15$, $p < .001$, and perspective, $F_{robust}(1, 597) = 24.98$, $p < .001$, remained significant. The interaction, as in the results reported in the main manuscript, was not significant, $F_{robust}(1, 597) = .13$, $p = .716$. Levene's test for mind perception was not significant, $F(3, 597) = .81$, $p = .491$. These results taken together, also noting the relatively large sample size and balanced design, indicate that heteroscedasticity is not a major problem in interpreting our results.

We also tested the possible effects of counterbalancing. A Levene's test including team type (IV), perspective (PERS), and counterbalancing (CNTRBLNC) indicated substantial heterogeneity of variance across the eight factorial cells, $F(7, 593) = 8.41$, $p < .001$. We therefore estimated a 2 (IV) × 2 (PERS) × 2 (CNTRBLNC) ANOVA on DV using HC3 heteroskedasticity-robust (White-corrected, Type III) standard errors. The robust ANOVA showed significant main effects of team type, $F_{robust}(1, 593) = 28.40$, $p < .001$, $p\eta^2 = .05$, and perspective, $F_{robust}(1, 593) = 24.95$, $p < .001$, $p\eta^2 = .04$. In contrast, the main effect of counterbalancing was not significant, $F_{robust}(1, 593) = 0.21$, $p = .65$, $p\eta^2 \approx .00$. Moreover, counterbalancing did not interact with team type or perspective, nor was the three-way interaction significant, all $F_{robust}(1, 593) \leq 1.00$, all $p \geq .32$, $p\eta^2 \approx .00$.

Repeating the counterbalancing analysis for mind perception showed no heteroskedasticity, $F(7, 593) = 1.11$, $p = .354$. We therefore estimated a standard 2 (IV) × 2 (PERS) × 2 (CNTRBLNC) Type III ANOVA on mind perception. The ANOVA showed a strong main effect of IV, $F(1, 593) = 145.30$, $p < .001$, $p\eta^2 = .20$. The main effects of PERS and counterbalancing were not significant, $F(1, 593) = 2.73$, $p = .099$, $p\eta^2 = .01$, and $F(1, 593) = .49$, $p = .48$, $p\eta^2 \approx .00$, respectively. However, IV interacted with PERS, $F(1, 593) = 3.89$, $p = .049$, $p\eta^2 = .01$, while the remaining interactions (including the three-way) were not significant, all $F(1, 593) \leq 1.28$, all $p \geq .26$, $p\eta^2 \approx .00$.

Including counterbalancing as a moderator in the PROCESS analysis (Model 11; 10,000 bootstrap resamples) did not meaningfully change the moderated mediation pattern reported in the main manuscript, either. In the first-stage model predicting mind perception, the three-way interaction (IV × PERS × CNTRBLNC) was not significant ($p = .258$). In the outcome model, mind perception again did not significantly predict responsibility when controlling for team type ($b$ = 1.05, $SE$ = 0.76, $p$ = .168), whereas the direct effect of team type on responsibility remained significant ($b$ = 11.00, $SE$ = 2.08, $p$ < .001). Accordingly, the conditional indirect effects of team type on responsibility through mind perception were not significant at any combination of perspective and counterbalancing (all 95% CIs included zero; effects ranged from $b$ = −1.63 to −0.97). Consistent with this, the index of moderated moderated mediation was not significant (*IMMM* = −.48, *BootSE* = .64, 95% CI [−1.97, .58]), and the indices of conditional moderated mediation by perspective were also not significant at either level of counterbalancing (CNTRBLNC = 0: *ICMM* = 0.65, *BootSE* = .65, 95% CI [−.41, 2.15]; CNTRBLNC = 1: *ICMM* = 0.18, *BootSE* = .41, 95% CI [−.47, 1.23]). Thus, the ordering manipulation did not meaningfully influence the conclusion that mind perception did not mediate (nor differentially mediate across perspectives) the effect of team type on responsibility. In conclusion, all our results are robust with regards to counterbalancing.

**References for the Web Appendix**